%% file: SA26-tech_comms.tex
\documentclass[sigconf]{acmart}
\usepackage{booktabs}
\usepackage{graphicx}
\usepackage{amsmath}
\usepackage{multirow}

\copyrightyear{2026}
\acmYear{2026}
\setcopyright{cc}
\setcctype{by}
\acmConference[SA Technical Communications '26]{SIGGRAPH Asia 2026 Technical Communications}{December 01--04, 2026}{Kuala Lumpur, Malaysia}
\acmBooktitle{SIGGRAPH Asia 2026 Technical Communications (SA Technical Communications '26), December 01--04, 2026, Kuala Lumpur, Malaysia}
\acmDOI{10.1145/3829339.3847861}
\acmISBN{979-8-4007-2841-9/2026/12}

\title{When Visual Quality Misleads: Intent Recognition under Rendered Avatar Distortions}
\author{Ning-Hsuan Chang}
\email{ninghsuan1818@gmail.com}
\orcid{0009-0006-9878-9266}
\affiliation{%
  \institution{National Chengchi University}
  \country{Taiwan}
}

\author{Kai-Siang Ma}
\email{seanmamasde@gmail.com}
\orcid{0009-0007-7958-2822}
\author{Yu-Chih Chen}
\email{berriechen@nycu.edu.tw}
\orcid{0000-0003-2710-083X}
\affiliation{%
  \institution{National Yang Ming Chiao Tung University}
  \country{Taiwan}
}
\thanks{This work was supported by the Higher Education Sprout Project of National Yang Ming Chiao Tung University (NYCU) and the Ministry of Education (MOE), Taiwan, under Grant 115W010161, and by the National Science and Technology Council (NSTC), Taiwan, under Grant NSTC 115-2813-C-A49-146-E. The human-subject study was approved by the Institutional Review Board of NYCU (IRB No. NYCU114137BE)}

\newif\ifshowrerun 
\showrerunfalse

\begin{document}

\begin{abstract}
Avatar-streaming systems are commonly evaluated with image and video quality assessment (IQA/VQA) metrics, implicitly treating visual fidelity as a proxy for communicative success. We test this assumption through a controlled behavioral study of rendered 3D avatars across a pristine condition and fourteen geometric, photometric, temporal, and combined distortions. Fifty-nine participants contributed 2,688 judgments of perceived action, response confidence, and visual quality. We identify \emph{Misleading Quality} in this dataset as distorted renderings that retain above-average perceived quality but yield below-average action-recognition accuracy. We also derive an Intent Quality Score (IQS) combining recognition correctness and confidence as the behavioral target for objective metrics. Among 126 distorted content--condition cells, 31 (24.6\%) exhibited Misleading Quality; temporal and geometric distortions showed the highest rates, at 50.0\% and 31.1\%, respectively. The results reveal a quality--accuracy dissociation where distortion families affect appearance and communication differently. Across 24 direct-scoring IQA/VQA metrics and three supervised feature-regression baselines, alignment with IQS remained limited; at $\lambda=0.5$, the best leave-one-content-out baseline reached PLCC $=0.4435$. Under this controlled protocol, visual fidelity alone is insufficient for avatar communication, motivating intent-aware quality assessment and streaming objectives.
\end{abstract}



\begin{teaserfigure}
  \centering
  \includegraphics[width=0.75\linewidth]{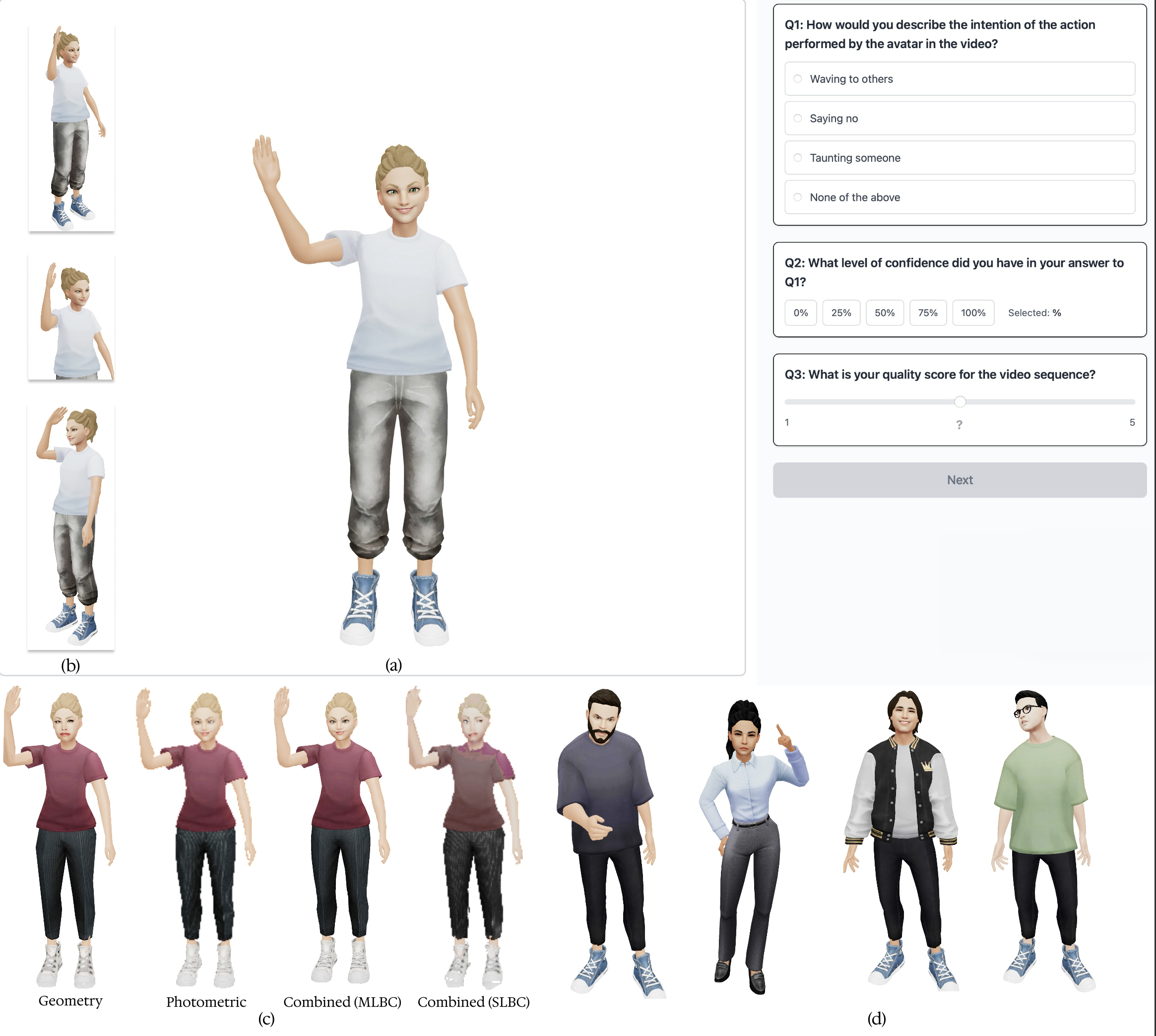}
  \caption{
    Overview of the behavioral benchmark.
    (a) Web interface for avatar inspection and reporting intent, confidence, and visual quality.
    (b) Free-orbit viewpoints.
    (c) Geometric, photometric, and combined distortions.
    (d) Avatar--action content units.
    Avatar assets \copyright~Ready Player Me O\"U; animations \copyright~Adobe Inc. (Mixamo).
  }
  \Description{
    A behavioral-study overview showing an avatar viewed from multiple angles, the web interface used to report action, confidence, and visual quality, examples of rendering distortions, and several avatar action contents.
  }
  \label{fig:platform_interface}
\end{teaserfigure}

\maketitle
\section{Introduction}

Avatar-streaming systems must balance visual fidelity against bandwidth and rendering cost and are therefore commonly optimized using image and video quality assessment (IQA/VQA) metrics~\cite{wang2023hermes,zhang2022yuzu}.
In communication settings, however, an avatar succeeds only when observers correctly understand its intended action or gesture
~\cite{cui2023virtualhuman,weidner2023avatar}. Geometric simplification, texture degradation, and frame-rate reduction may leave the overall rendering visually acceptable while selectively corrupting pose, motion, or geometric cues that carry communicative meaning. Visual fidelity and communicative success are therefore related but distinct evaluation targets.

Prior work on rendered avatars, point clouds, and dynamic meshes has established subjective quality datasets and objective quality predictors
~\cite{nehme2021visual,chen2024avatar,cui2023sjtu}. Studies of graphics perception show that observers are sensitive to changes in
character motion, pose-update rate, and gesture parameters
~\cite{reitsma2003perceptual,ferstl2020gesture}, while task-oriented video-quality research has examined whether compression preserves
intelligibility, notably for sign-language communication
~\cite{cavender2008mobileasl,pinson2013task}. 
These studies leave open whether conventional IQA/VQA metrics can predict whether observers recover a rendered avatar's intended action.

We address this gap through a controlled behavioral study in which 59 participants provided 2,688 judgments on ten avatar--action content units rendered under one pristine condition and fourteen distortions spanning geometric, photometric, temporal, and combined degradation families. For each stimulus, participants reported the perceived action, response confidence, and visual quality. From recognition correctness and confidence, we derive an Intent Quality Score (IQS) for metric evaluation. Joint analysis of perceived quality and recognition accuracy allows us to characterize \emph{Misleading Quality}: renderings that remain visually acceptable yet fail to preserve communicative intent. Temporal and geometric distortions produce the largest proportions of such cases, while 24 direct-scoring quality metrics and three supervised
feature-regression baselines show limited correspondence with IQS; the strongest reaches PLCC $=0.4435$ under leave-one-content-out validation. Thus, visual quality alone is insufficient for evaluating rendered-avatar communication.

We contribute a controlled behavioral protocol, an operational characterization of Misleading Quality, and a benchmark showing the limited intent sensitivity of existing IQA/VQA methods.

\section{Controlled Study and Evaluation Protocol}

\paragraph{Stimuli and behavioral study.}
We used five stylized avatars~\cite{readyplayerme}, each paired with
two distinct skeletal animations~\cite{mixamo}, yielding ten avatar--action content units. Each unit was rendered under fifteen conditions---fourteen distortions and one pristine baseline
(ORIG)---for 150 stimuli. The distortions span geometric, photometric, temporal, and combined families and serve as controlled rendering-level approximations rather than outputs of specific
production codecs.

Fifty-nine participants freely orbited and inspected randomized stimulus subsets through a web-based Three.js interface, contributing 2,688 judgments (approximately 18 per stimulus). For each clip, they selected the perceived action from three to five pilot-tested, semantically similar options, rated confidence on a five-level scale, and rated visual quality on a five-point absolute category scale~\cite{itur_bt500}. The study received institutional review-board
approval, and all participants provided informed consent. We required pristine-condition response consensus of at least $0.6$: E5 was excluded ($0.5556$), whereas E6 was retained
($0.6111$). The analyses therefore use nine contents (E1--E4 and E6--E10), comprising 135 content--condition cells, of which 126 are
distorted.

\paragraph{Behavioral target.}
We report cell-level recognition accuracy and mean opinion score (MOS). For metric evaluation, each response is represented by the pragmatic confidence-weighted task score
\begin{equation}
    \mathrm{IQS}=ac-\lambda(1-a)c,
    \label{eq:iqs}
\end{equation}
where $a\in\{0,1\}$ denotes recognition correctness, $c\in[0,1]$ is normalized confidence, and $\lambda$ penalizes confident
misidentification. IQS is averaged within each content--condition cell. Main results use $\lambda=0.5$; supplementary analysis confirms
the same conclusions for $\lambda\in\{0,0.25,0.5,0.75,1.0\}$.

\paragraph{Objective metric evaluation.}
We evaluated 27 methods: 24 direct-scoring metrics spanning FR-IQA, NR-IQA, FR-VQA, and NR-VQA, together with three supervised baselines built on CONTRIQUE~\cite{madhusudana2022conrique}, Re-IQA~\cite{saha2023reiqa}, and DreamSim~\cite{fu2023dreamsim} features. A deterministic Three.js--Playwright pipeline produced $1024\times1024$ still images
from a fixed front-oriented canonical view at animation time $1.0$~s and ten-frame $512\times512$ videos over $0.50$--$1.50$~s, encoded as 1-s, 10-fps H.264 clips. Full-reference methods used the corresponding ORIG rendering, and the two low-frame-rate conditions were excluded from still-image analyses. This yielded 117 NR-IQA cells, 108 FR-IQA distorted--reference pairs, 135 NR-VQA cells, and 126 FR-VQA
distorted--reference pairs. Feature regressors used fixed-seed leave-one-content-out validation, holding out one avatar--action unit. Because participants freely orbited the
avatars while metrics used one canonical view, the correlations characterize
this controlled protocol rather than unrestricted interactive or HMD-based
perception. Full implementation settings are in the supplementary
material.

\section{Results}

\begin{figure*}[t]
  \centering
  \includegraphics[width=0.97\linewidth]{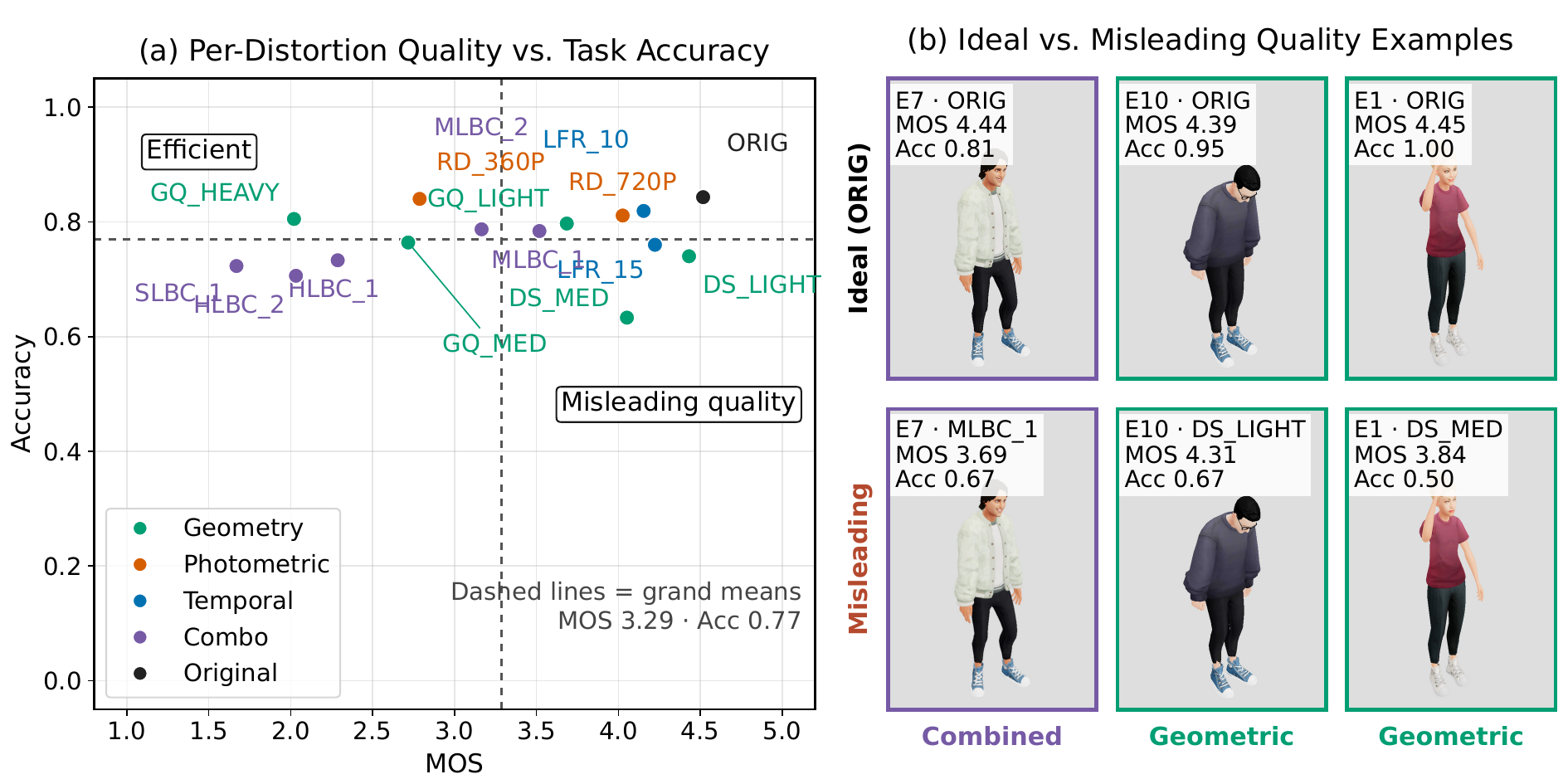}
  \caption{\emph{Misleading Quality} and quality--accuracy dissociation. 
    (a) MOS and recognition accuracy by distortion. Dashed lines show grand means (MOS $=3.29$, accuracy $=0.77$). Lower-right conditions are visually acceptable
    but behaviorally unsuccessful; upper-left conditions show the reverse.
    (b) Pristine and misleading-quality examples. Prevalence uses
    content--condition cells with these thresholds.
  }
  \Description{
    A scatter plot compares mean visual-quality scores with action-recognition accuracy across distortion conditions. Dashed lines divide the plot at the grand-mean MOS of 3.29 and accuracy of 0.77. Several conditions fall in the lower-right misleading-quality region, where visual quality remains relatively high while recognition accuracy is low. Example pristine and misleading avatar renderings are shown beside the plot.
  }
  \label{fig:four_quadrant_with_examples}
\end{figure*}

\textbf{Misleading Quality and quality--accuracy dissociation.} We define a distorted content--condition cell as exhibiting \emph{Misleading Quality} when its MOS is above the grand mean across all analyzed conditions ($3.29$) while its recognition accuracy is below the corresponding grand mean ($0.77$). These dataset-based thresholds provide a practical way to identify quality--recognition divergence. Figure~\ref{fig:four_quadrant_with_examples}(a) shows the same thresholds using per-distortion averages, while Fig.~\ref{fig:four_quadrant_with_examples}(b) compares pristine and distorted renderings that remain visually acceptable but reduce action recognition. Unlike visibly degraded renderings, these cases give little indication of lost communication-relevant information, so they may appear satisfactory even when the intended action is not reliably conveyed.

Across 126 distorted content--condition cells, 31 (24.6\%) fell into the misleading-quality quadrant. Temporal low-frame-rate distortions had the highest prevalence (9/18, 50.0\%), followed by geometric distortions (14/45, 31.1\%). Photometric and combined rates were 3/18 (16.7\%) and 5/45 (11.1\%), respectively. The opposite upper-left quadrant further demonstrates the quality--accuracy dissociation: some visually salient degradations substantially reduce MOS while preserving recognition, whereas geometric or temporal degradations can impair communication without a proportional loss in perceived quality. Visual degradation and communicative failure are therefore neither equivalent nor monotonically coupled.

\textbf{Existing metrics weakly predict communicative intent.} Across 24 direct-scoring IQA/VQA metrics, no method exceeded $|\mathrm{PLCC}|=0.3237$ over $\lambda\in\{0,0.25,0.5,0.75,1\}$. Table~\ref{tab:regression_reliability} summarizes three feature-regression baselines at $\lambda=0.5$. Re-IQA was strongest, reaching PLCC $=0.4435$ and SRCC $=0.4501$ under leave-one-content-out validation. Confidence intervals remain broad; several for CONTRIQUE and DreamSim include zero. Thus, fine-grained differences are descriptive rather than reliable rankings.

Split-half IQS agreement reached PLCC $=0.6381$, above all objective methods. IQS is reproducible, though not noise-free, and target unreliability alone cannot explain limited metric performance.

\begin{table*}[t]
\centering
\caption{Leave-one-content-out feature-regression performance against IQS and split-half IQS agreement at $\lambda=0.5$. Confidence intervals are 95\% content-block bootstrap intervals.}
\Description{
  The table reports PLCC, SRCC, and KRCC correlations with 95 percent confidence intervals for CONTRIQUE, Re-IQA, DreamSim, and split-half IQS agreement. Re-IQA has the highest correlations among the three feature-regression baselines, while split-half IQS agreement is higher than all evaluated baselines.
}
\label{tab:regression_reliability}
\footnotesize
\setlength{\tabcolsep}{3pt}
\renewcommand{\arraystretch}{1.12}
\begin{tabular}{@{}lcccccc@{}}
\toprule
\textbf{Evaluation} &
\textbf{PLCC} &
\textbf{95\% CI} &
\textbf{SRCC} &
\textbf{95\% CI} &
\textbf{KRCC} &
\textbf{95\% CI} \\
\midrule
\multicolumn{7}{l}{\textit{Feature-regression baselines}} \\
CONTRIQUE &
0.2923 &
[$-0.0193$, 0.5627] &
0.2742 &
[$-0.0117$, 0.5543] &
0.1871 &
[$-0.0076$, 0.4087] \\
Re-IQA &
0.4435 &
[0.1260, 0.6206] &
0.4501 &
[0.0403, 0.6800] &
0.3135 &
[0.0237, 0.4971] \\
DreamSim &
0.3795 &
[$-0.1298$, 0.6242] &
0.3470 &
[$-0.1204$, 0.6361] &
0.2364 &
[$-0.0673$, 0.4628] \\
\midrule
\multicolumn{7}{l}{\textit{Behavioral reliability}} \\
IQS split-half agreement &
0.6381 &
[0.5562, 0.7147] &
0.6444 &
[0.5658, 0.7182] &
0.4662 &
[0.4026, 0.5312] \\
\bottomrule
\end{tabular}
\end{table*}

\section{Discussion, Limitations, and Conclusion}

The results indicate that visual fidelity and communicative effectiveness should be treated as complementary rather than interchangeable objectives in rendered-avatar streaming. Conventional bitrate-allocation and level-of-detail strategies primarily optimize perceptual appearance~\cite{schwarz2018mpeg,zhang2022yuzu}, yet under the tested distortions, temporal and geometric degradations more often reduced action recognition even when perceived quality remained relatively high. Improving an IQA/VQA score therefore does not necessarily protect the motion, pose, or geometric cues that support intent recognition. A practical direction is to augment conventional perceptual objectives with intent-aware evaluation or content-adaptive protection of communication-relevant cues. This study motivates such an objective but does not yet provide a deployable task-aware quality metric.

These conclusions are restricted to the evaluated protocol. Participants viewed monoscopic desktop renderings with free-orbit inspection rather than stereoscopic HMD content with embodied interaction~\cite{weidner2023avatar,chen2024avatar}. Conversely, objective metrics were evaluated using fixed canonical-view images or videos, so the reported correlations do not characterize view-aware or interactive quality models. The evaluated distortions are controlled rendering-level approximations rather than artifacts generated by specific production codecs. The analysis uses nine retained avatar--action units from five stylized avatars, with intent assessed through constrained response choices. Because each animation was paired with a single avatar, action and character effects cannot be disentangled. Moreover, distortion levels were not perceptually matched across families; family-level Misleading Quality rates are therefore descriptive of the tested conditions rather than a causal ranking of distortion importance. IQS is also a pragmatic confidence-weighted target rather than a complete model of communicative understanding. Broader validation should therefore examine HMD-based viewing, codec-generated artifacts, factorial combinations of avatars and actions, perceptually matched distortion severities, more diverse avatar styles and communicative behaviors, and metrics that jointly model viewpoint, motion, and task relevance. Within these constraints, better-looking avatars do not necessarily communicate better.

\bibliographystyle{ACM-Reference-Format}
\bibliography{references}

\input{SA26-tech_comms_supp}

\end{document}

%% file: SA26-tech_comms_supp.tex
\clearpage
\appendix
\twocolumn[
\begin{@twocolumnfalse}
    \begin{center}
        \Huge \textbf{Appendix}
    \end{center}
    \vspace{0.5cm}
\end{@twocolumnfalse}
]
\begin{figure*}
\centering
\includegraphics[width=0.88\linewidth,trim=0 0 0 24,clip]{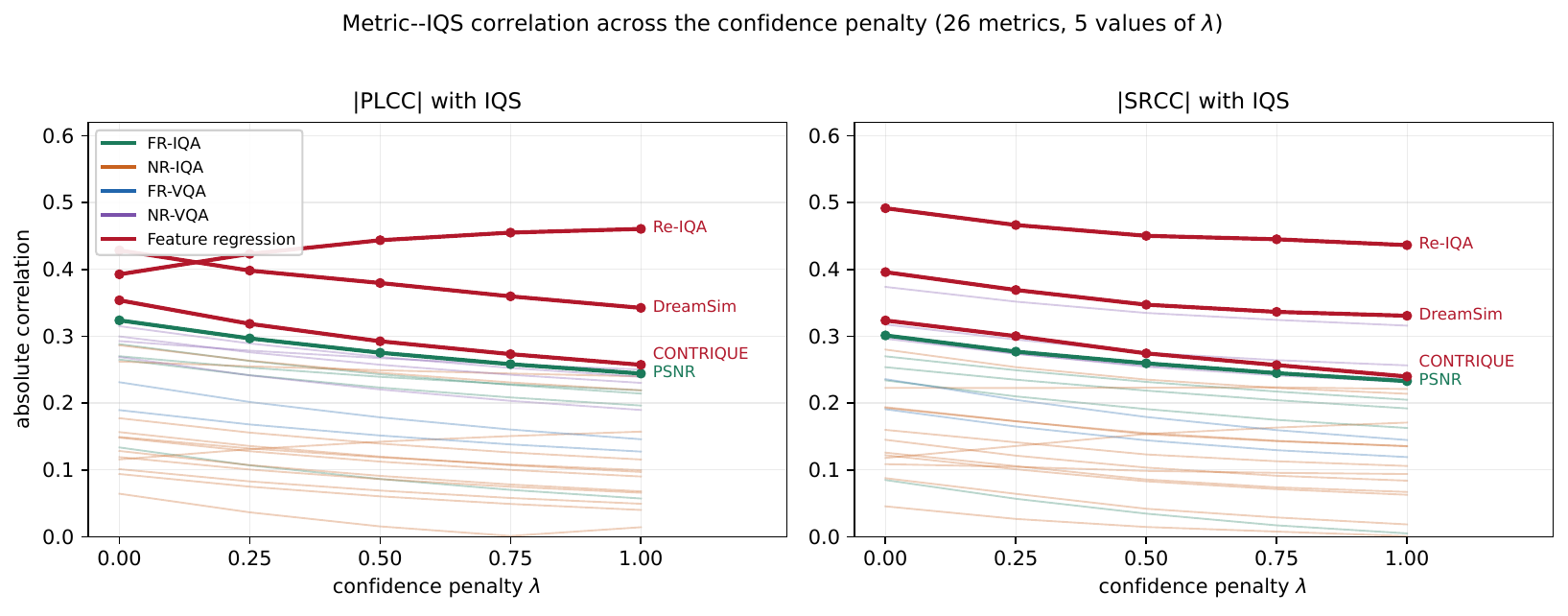}
\caption{Sensitivity of metric--IQS correspondence to the confidence penalty $\lambda$. Curves show absolute PLCC and SRCC for 24 direct-scoring metrics and three feature-regression baselines. Emphasized curves identify the strongest methods near $\lambda=0.5$, while the remaining direct-scoring metrics are shown faintly by family.}
\Description{
Two plots show absolute PLCC and SRCC between IQS and the evaluated metrics
as the confidence penalty $\lambda$ varies from 0 to 1.
Curves represent 24 direct-scoring metrics and three feature-regression
baselines, with the strongest methods emphasized.
}
\label{fig:lambda_sensitivity}
\end{figure*}

\section{Stimuli, Distortions, and Evaluation Details}

This supplementary material provides additional implementation details for the ten avatar--action content units and fifteen rendering conditions summarized in the main paper. The fourteen distortions were implemented through Three.js/WebGL operations to isolate geometric, photometric, temporal, and combined degradation characteristics. They should be interpreted as controlled rendering-level approximations rather than outputs of production codecs such as Draco or V-PCC.

\subsection{Avatar Assets and Actions}

We used five stylized 3D avatar characters generated with Ready Player Me, each paired with two motion-capture animations from the Mixamo library, yielding ten avatar--action content units (E1--E10). The action set covers greeting, affective, and socially communicative behaviors. Ground-truth action labels were determined from the modal response under the pristine ORIG condition. Actions were retained only when the pristine-condition consensus ratio reached at least $0.6$; E5 was excluded, whereas E6 was retained.

\begin{table}[t]
\centering
\caption{Avatar--action content units used in the study.}
\Description{
Ten avatar--action content units and their action descriptions.
E5 was excluded from analysis because its pristine-condition
consensus ratio was below the required threshold.
}
\label{tab:action_categories}
\footnotesize
\setlength{\tabcolsep}{3pt}
\renewcommand{\arraystretch}{1.05}
\begin{tabular}{@{}clp{4.8cm}@{}}
\toprule
\textbf{ID} & \textbf{Action} & \textbf{Description} \\
\midrule
E1 & Waving & Greeting gesture with raised hand and outward-facing palm \\
E2 & Thumbs-up & Positive acknowledgment with the thumb extended upward \\
E3 & Angry yell & Aggressive expression with open mouth and tense posture \\
E4 & Angry point & Accusatory gesture with an extended arm and finger \\
E5 & Crazy gesture$^*$ & Erratic arm movements \\
E6 & Begging & Pleading posture with clasped hands and lowered stance \\
E7 & Agreement & Nodding with an affirmative hand gesture \\
E8 & Shaking fist & Threatening gesture with a clenched hand \\
E9 & Scornful look & Disdainful expression with a turned head and eye roll \\
E10 & Thankful bow & Gratitude expressed through a forward-leaning posture \\
\bottomrule
\end{tabular}
\vspace{2pt}

{\footnotesize $^*$Excluded from analysis because its pristine-condition consensus ratio was $0.5556<0.6$.}
\end{table}

\input{new_direct_full}

\begin{table*}[t]
\centering
\caption{Summary of the controlled distortion families and condition labels. Combined conditions were implemented as fixed presets of geometric, photometric, and/or temporal operations.}
\Description{
Controlled distortion conditions grouped into geometric, photometric, temporal, and combined families, with the rendering operation used for each condition.
}
\label{tab:distortion_settings}
\footnotesize
\setlength{\tabcolsep}{3.5pt}
\begin{tabular}{@{}lll@{}}
\toprule
\textbf{Family} & \textbf{Condition} & \textbf{Generation setting} \\
\midrule
Geometry & GQ\_LIGHT, GQ\_MED, GQ\_HEAVY & Vertex-position quantization with $\delta\in\{0.004,0.008,0.016\}$ \\
Geometry & DS\_LIGHT, DS\_MED & Mesh downsampling to 50\% and 25\% of the original geometry \\
Photometric & RD\_720P, RD\_360P & Spatial-resolution reduction to 720p and 360p \\
Temporal & LFR\_15, LFR\_10 & Frame-rate reduction to 15 and 10 fps \\
Combined & MLBC\_1, MLBC\_2, HLBC\_1, HLBC\_2, SLBC\_1 & Fixed multi-operation presets used in the stimulus set \\
\bottomrule
\end{tabular}
\end{table*}

\subsection{Distortion Conditions}
The combined presets were treated as named experimental conditions. Their constituent operations were not varied factorially, and the study therefore does not infer component-level effects from comparisons among these presets.

Fifty-nine participants evaluated randomized stimulus subsets and contributed 2,688 valid judgments, corresponding to approximately 18 ratings per stimulus. For each clip, participants selected the perceived action from three to five action-specific, semantically similar options validated through pilot testing, rated confidence on a five-level scale, and rated visual quality on a five-point absolute category scale. The response alternatives were designed to represent plausible fine-grained action confusions rather than trivially unrelated choices. Pristine-condition response consensus was required to be at least $0.6$: E5 was excluded because its consensus was $0.5556$, whereas E6 was retained at $0.6111$. The retained dataset therefore contains nine contents (E1--E4 and E6--E10), 135 content--condition cells including ORIG, and 126 distorted cells. Because the action-specific response sets contained three to five options, raw recognition accuracy was not perfectly chance-equated across actions; nevertheless, the same observed cell-level accuracy definition was applied consistently across all conditions.

A deterministic rendering pipeline based on Three.js and Playwright generated all objective-metric inputs. Still images were rendered as $1024\times1024$ PNGs at animation time $1.0$~s. A $50^{\circ}$ perspective camera targeted the center of the animated mesh bounding box from direction $(-2,2,3)$. Camera distance was adapted to content according to
\begin{equation}
d=\frac{1.6m}{2\tan(25^{\circ})},
\label{eq}
\end{equation}
where $m$ denotes the largest bounding-box dimension. Each scene used a uniform light-gray background, sRGB output, ACES tone mapping, one ambient light, and three directional lights. Full-reference image metrics used the corresponding ORIG avatar--action rendering. The LFR\_15 and LFR\_10 conditions were excluded from still-image analyses because their degradation is temporal rather than spatial.

Video inputs contained ten $512\times512$ frames sampled over $0.50$--$1.50$~s and encoded as 1-s, 10-fps H.264 MP4 clips. Full-reference video metrics used temporally aligned ORIG clips. This protocol yielded 117 NR-IQA cells, 108 FR-IQA distorted--reference pairs, 135 NR-VQA cells, and 126 FR-VQA distorted--reference pairs.

CONTRIQUE, Re-IQA, and DreamSim were additionally evaluated as fixed feature representations followed by supervised regression. Fixed-seed leave-one-content-out (LOCO) validation held out all cells associated with one avatar--action content unit, thereby preventing interpolation between distorted versions of the same content across the training and test sets.

\section{IQS Robustness and Statistical Reliability}

IQS is used as a pragmatic confidence-weighted task score rather than as a formal signal-detection-theory measure. For response $i$, it is defined as
\begin{equation}
\mathrm{IQS}_{i}(\lambda)=a_{i}c_{i}-\lambda(1-a_{i})c_{i},
\label{eq:iqs_supp}
\end{equation}
where $a_{i}\in\{0,1\}$ denotes recognition correctness and $c_{i}\in[0,1]$ denotes normalized response confidence. At $\lambda=0$, incorrect responses receive a score of zero; increasing $\lambda$ assigns a progressively larger penalty to confident misidentification. Sensitivity was evaluated over $\lambda\in\{0,0.25,0.5,0.75,1\}$.

Across the full sensitivity sweep, the strongest feature-regression result reached $|\mathrm{PLCC}|=0.4604$, whereas no direct-scoring metric exceeded $|\mathrm{PLCC}|=0.3237$ (Table~\ref{tab:direct_rerun}). Re-IQA was the strongest feature representation for $\lambda\geq0.25$, while DreamSim was marginally stronger at $\lambda=0$. PSNR remained the strongest direct FR-IQA method throughout the sweep. Within NR-VQA, VSFA performed best for $\lambda\leq0.5$, while DOVER-Mobile performed best for $\lambda\geq0.75$; their PLCC difference remained below $0.02$ at every tested value. These sensitivity results indicate that the limited association between existing metrics and IQS is not specific to the primary choice of $\lambda=0.5$. Absolute correlations are reported in this analysis because the native score directions differ across metrics and the purpose of the sweep is to compare association magnitude rather than score polarity.

Correlation uncertainty was quantified using 10,000 content-block bootstrap resamples with fixed random seed 20260727. In each resample, the nine retained content units were sampled with replacement, and all content--condition cells belonging to a sampled content unit remained grouped when the correlations were recomputed. This procedure preserves within-content dependence across distortion conditions and avoids treating individual cells as independent observations.

Split-half agreement was estimated at the response level within each content--condition cell. Ratings in each cell were randomly divided into two groups, cell-level IQS was recomputed independently for the two halves, and the resulting cell vectors were correlated. The corresponding feature-regression results and split-half agreement at $\lambda=0.5$ are reported in Table~\ref{tab:regression_reliability} of the main paper. Split-half agreement is used as a behavioral reliability reference rather than as a literal upper bound on the performance of every possible predictor.

%% file: new_direct_full.tex
\begin{table}[t]
\centering
\caption{Direct-scoring metric correlations with IQS by metric family. Brackets denote 95\% content-block bootstrap confidence intervals; rows are ordered by PLCC within each family.}
\Description{
Direct-scoring metric correlations with IQS, grouped into FR-IQA, NR-IQA,
FR-VQA, and NR-VQA families. The table reports PLCC values and 95 percent
content-block bootstrap confidence intervals for each metric.
}
\label{tab:direct_rerun}
\footnotesize
\setlength{\tabcolsep}{4pt}
\renewcommand{\arraystretch}{1.05}
\begin{tabular}{@{}llcc@{}}
\toprule
\textbf{Family} & \textbf{Metric} & \textbf{PLCC} & \textbf{95\% CI} \\
\midrule
\multirow{5}{*}{FR-IQA}
& PSNR & 0.2752 & [0.090, 0.452] \\
& FSIM & 0.2429 & [0.106, 0.412] \\
& SSIM & 0.2392 & [0.123, 0.384] \\
& VIF & 0.2231 & [0.142, 0.306] \\
& LPIPS & 0.0863 & [$-0.017$, 0.219] \\
\midrule
\multirow{12}{*}{NR-IQA}
& NIQE & 0.2491 & [$-0.085$, 0.434] \\
& MUSIQ & 0.2453 & [0.076, 0.375] \\
& PaQ-2-PiQ & 0.1392 & [0.012, 0.250] \\
& TReS & 0.1198 & [$-0.002$, 0.236] \\
& MANIQA & 0.1187 & [$-0.034$, 0.277] \\
& LIQE & 0.1123 & [0.005, 0.233] \\
& DBCNN & 0.0910 & [$-0.022$, 0.199] \\
& TOPIQ\_NR & 0.0861 & [$-0.017$, 0.199] \\
& BRISQUE & 0.0690 & [$-0.217$, 0.274] \\
& CLIPIQA+ & 0.0603 & [$-0.072$, 0.275] \\
& ARNIQA & 0.0154 & [$-0.165$, 0.204] \\
& NIMA & $-0.1421$ & [$-0.344$, 0.228] \\
\midrule
\multirow{2}{*}{FR-VQA}
& ST-RRED & 0.1785 & [0.078, 0.325] \\
& VMAF & 0.1514 & [0.080, 0.220] \\
\midrule
\multirow{4}{*}{NR-VQA}
& VSFA & 0.2689 & [0.032, 0.470] \\
& DOVER-Mobile & 0.2671 & [$-0.092$, 0.509] \\
& DOVER & 0.2572 & [0.046, 0.439] \\
& FAST-VQA & 0.2202 & [0.029, 0.425] \\
\bottomrule
\end{tabular}
\end{table}